\documentclass[aps,prb,amsmath,amssymb,twocolumn]{revtex4-2}
\usepackage{graphicx}
\usepackage{bm}
\usepackage{mathptmx}
\usepackage{epstopdf}
\usepackage{booktabs}
\usepackage{multirow}
\usepackage{hhline}
\usepackage{subfigure}
\usepackage{tabularx}
\usepackage{float}
\begin{document}
\newcommand{\ms}{m_*}
\newcommand{\bor}{{\bf r}}
\newcommand{\mm}{{\bf m}}
\newcommand{\rdot}{\dot{\bf r}}
\newcommand{\rddot}{\dot\dot{\bf r}}
\newcommand{\es}{\bf{E^s_\sigma}}

\title{Customizing PBE Exchange-Correlation functionals: A comprehensive approach for band gap prediction in diverse semiconductors}
\author{Satadeep Bhattacharjee}
\email{s.bhattacharjee@ikst.res.in}
\author{Namitha Anna Koshi}
\affiliation{Indo-Korea Science and Technology Center (IKST), Bangalore, India}
\author{Seung-Cheol Lee}
\affiliation{Electronic Materials Research Center, Korea Institute of Science $\&$ Technology, Korea}
\begin{abstract} 
Accurate band gap prediction in semiconductors is crucial for materials science and semiconductor technology advancements. This paper extends the Perdew-Burke-Ernzerhof (PBE) functional for a wide range of semiconductors, tackling the exchange and correlation enhancement factor complexities within Density Functional Theory (DFT). Our customized functionals offer a clearer and more realistic alternative to DFT+U methods, which demand large negative U values for elements like Sulfur (S), Selenium (Se), and Phosphorus (P). Moreover, these functionals are more cost-effective than GW or Heyd-Scuseria-Ernzerhof (HSE) hybrid functional methods, therefore, significantly facilitating the way for unified workflows in analyzing electronic structure, dielectric constants, effective masses, and further transport and elastic properties, allowing for seamless calculations across various properties. We point out that such development could be helpful in the creation of comprehensive databases of band gap and dielectric properties of the materials without expensive calculations. Furthermore, for the semiconductors studied, we show that these customized functionals and the Strongly Constrained and Appropriately Normed semilocal density functional (SCAN) perform similarly in terms of the band gap.

\end{abstract}
\keywords{Exchange-correlation, band gap}
\maketitle
\section{Introduction}
The failure of standard density functional theory (DFT) based calculations to predict the correct band gap of semiconductors is well-known in solid-state physics and materials science~\cite{problem1}. This failure comes from an inappropriate interpretation of the Kohn-Sham eigenvalues. A key issue is the lack of derivative discontinuity in the total energy relative to occupation numbers at integer values~\cite{dis}, leading to a notable underestimation of band gaps~\cite{gap1,gap2}. The \textit{exchange and correlation (XC) functional}, which is approximated because the precise XC functional is still unknown, is a crucial component of DFT, despite being an exact theory. 
To mitigate these issues, various solutions have been proposed. One such remedy is the employment of hybrid functionals~\cite{hybrid}, which merge LDA (local density approximation) or GGA (generalized gradient approximation) with a portion of exact exchange, thereby enhancing band gap prediction accuracy. Another strategy is the GW method, a sophisticated many-body perturbation theory that computes the system's self-energy to deliver a more precise depiction of electronic structures and band gaps. However, both these methods are computationally intensive and resource-demanding. Furthermore, integrating them into a workflow for calculating post-processing aspects, such as electronic and thermal transports, is challenging. A more straightforward and accessible alternative is the DFT+U method~\cite{anisimov1991band,liechtenstein1995density}, which has been effectively used to rectify inaccurate band gap predictions in semiconductors by standard LDA and GGA functionals. This method incorporates the Hubbard U term, adjustable as either positive or negative, to amend ground-state properties and band gap inaccuracies. Positive U values are instrumental in discouraging integer occupations and promoting localization, advantageous in scenarios where the DFT+U electronic state, although higher in energy, can achieve metastable convergence. Conversely, negative U values are beneficial in enhancing DFT+U calculations for materials with less pronounced localized bands. An unconventional use of DFT+U with a negative U on phosphorus p-orbital has been demonstrated to improve the results~\cite{n1,n2}.

Though DFT+U calculations give fairly reasonable value of band gap for many semiconductors at reduced computational cost, it requires seemingly unphysical U$_{s}$ and U$_{p}$ values for s and p-states respectively, in addition to U$_{d}$. For instance, in the case of w-ZnO, a gap of 3.3 eV close to the experimental value of 3.4 eV is obtained by Paudel and Lambrecht, by using U$_{s}$ = 43.54 eV and U$_{d}$ = 3.40 eV \cite{Paudel2008}. There are few reports on employing negative values of U$_{p}$ for S/Se/Te in Zn/Cd monochalcogenides \cite{Khan2023, Khan2019, Menendez2014, Andriotis2013, Baldissera2016, Persson2006}. It is argued that negative values of U are appropriate for delocalized states (such as s and p states), where the exchange-correlation hole is overestimated by GGA. When the chalcogen ion is O, the positive value of U$_{p}$ is used \cite{Ma2013}, which increases the on-site repulsive Coulomb potential, in contrast to negative U corresponding to an attractive Coulomb potential. For metal oxides like NiO, CeO$_{2}$ and Ce$_{2}$O$_{3}$, the hybrid functionals which use a portion of Fock exchange yield band gaps in good agreement with experimental data \cite{Moreira2002, Hay2006}.  Uddin and Scuseria estimated the properties of four phases of bulk ZnO using screened exchange Heyd-Scuseria-Ernzerhof (HSE) functional and compared it with three other functionals (LDA, PBE formalism of GGA and Tao-Perdew-Staroverov-Scuseria meta-GGA) \cite{Uddin2006}. They have obtained improved electronic and optical properties, with the lattice constants, bulk moduli and cohesive energies agreeing well with experiments. The position of Zn 3d band, misrepresented in conventional DFT is better placed with a band gap of 2.9 eV for w-ZnO, which is 14.7\% less than the experimental value. 

The exchange-correlation functional plays a crucial role in describing electron-electron interactions in the field of DFT. Compared to the basic LDA, the GGA approach is an improvement. The electron density gradient is taken into account by GGA, whereas LDA only takes the local electron density into account. GGA therefore enables a more nuanced approach for cases where density fluctuates rapidly.

The inclusion of the gradient term in GGA aims to capture variations in electron density more effectively, especially in regions where density changes rapidly, such as near atomic nuclei or within molecular bond areas. However, determining the exact weight or emphasis for this gradient term in relation to the local term remains a topic of exploration.

To achieve closer alignment with experimental data or results from high-fidelity quantum mechanical methods, researchers frequently adjust these weights in the GGA functional. This optimization process involves fine-tuning the combination of local and gradient terms in the functional. Consequently, the GGA can be tailored to more accurately predict specific properties or behaviors of crystalline or molecular systems~\cite{xu2004extended,yu2016mn15,carmona2015generalized}.

In the present work, we propose two customized PBE ~\cite{perdew1996generalized} exchange-correlation functionals to effectively predict the semiconductor bandgap without large cost-effective calculations. Furthermore, they replace the two DFT+U approaches in semiconductors with both positive and negative U values. We refer to them as type-I and type-II functionals.
Especially compared to the approach that applies large negative $U$ values for S-p, Se-p etc. orbitals ( for O-p orbitals, positive U values are used in oxide semiconductors). This direct modification of the PBE function is not only a more transparent approach but also associated with more meaningful physics. Furthermore, in modern computational materials science, creating databases of material properties, like transport data, through HSE or GW methods is challenging due to computational demands. Researchers use a mix of precise and feasible methods, like HSE for electronic structure and PBE+U for dielectric constants, to manage this, though it sometimes compromises accuracy, especially in effective mass calculations. Our method offers accurate effective masses, enabling consistent application in electronic structure, dielectric constant determination, and transport property analysis, applicable to various material calculations. In the following section, we discuss the construction of the customized exchange-correlation functionals in detail.
\section{Theoretical framework: Customized exchange-correlation functionals}
The GGA-exchange correlation energy is given by,
\begin{equation}
\begin{split}
E_{GGA}^{XC} = & E_X[n(r)]+E_C[n(r)]\\ 
=&\int d^3r n(r) \epsilon_{\text{unif}}^x(n) F_{XC}(r_s, \xi, s)
\end{split}
\label{xc}
\end{equation}
Here $n(r)$ is the electron density. $\epsilon_{\text{unif}}^x(n) = -\frac{3k_f}{4\pi}$ is the Slater exchange energy of an uniform gas, $k_f = (3\pi^2 n)^{\frac{1}{3}}$ is the Fermi-wave vector. The factor $F_{XC}(r_s, \xi, s)$ is the so-called exchange-correlation enhancement factor which can further be expressed as the sum of the exchange part and the correlation part as, 
\begin{equation}
F_{XC}(r_s, \xi, s) = F_X(s) + \frac{\epsilon_{unif}^c(r_s, \xi)}{\epsilon_x(n)} F_C(r_s, \xi, t)
\label{enhance}
\end{equation}
Here $s = \frac{|\nabla n|}{2k_f n}$ is the dimensionless density gradient. $\xi = \frac{n^{\uparrow} - n^{\downarrow}}{n}$ is spin polarization, which is zero for the non-magnetic systems. 
$t = \frac{|\nabla n|}{(2\phi k_s n)}$ is a scaled density gradient and $\phi = [\left(1 + \zeta\right)^{2/3} + \left(1 - \zeta\right)^{2/3}]/2$
is a spin-scaling factor. $k_s = \left(4k_f/\pi\right)^{1/2}$ is the Thomas-Fermi screening
wave vector. The most popular GGA-exchange correlation is probably the PBE functional~\cite{ernzerhof1999assessment}. From atoms to solids, the PBE-functional is intended to be accurate for a broad variety of systems. It has been demonstrated to outperform LDA in numerous aspects, including band gaps, atomization energies, and bond lengths. It does, however, still have certain shortcomings and restrictions, including the inability to adequately characterize strongly correlated systems, the overestimation of molecular dissociation energies and the underestimation of van der Waals interactions. As a result, there are numerous additional GGA extensions and variants that seek to increase the precision and effectiveness of DFT calculations for various system kinds and property types~\cite{madsen2007functional,kraus2021extrapolating,Li2020Application,SCAN,Liu2023Implementation,Santra2020What}. In this work, we introduce refined functional forms for both $F_X(s)$ and $F_C(r_s,\xi,t)$, customized to more accurately represent exchange and correlation effects in semiconductors such as Zn-chalcogenides. This refinement addresses the shortcomings of standard DFT-functional based approaches, such as LDA and PBE, which often inaccurately estimate the exchange-correlation hole in these systems, leading to either significant overestimation or underestimation.

Within the framework of PBE exchange-correlation functional, the exchange enhancement factor $F_X(s)$ takes the form~\cite{perdew1996generalized},
\begin{equation}
F^{\text{PBE}}_X(s) = 1 + \kappa - \frac{\kappa}{(1 + \frac{\mu}{\kappa} s^2)}
\label{X-en}
\end{equation}
The parameter $\kappa=0.804$ was used to ensure the Lieb-Oxford bound $(E_X[n]\ge E_{XC}[n]\ge -1.68e^2\int d^3r n^{4/3}$) while the parameter $\mu=0.21951$ aims to achieve the precise cancellation of the second-order gradient expansion terms for exchange and correlation. The correlation part of the enhancement factor can be written as,
\begin{equation}
F_C(r_s, \xi, t) = 1 + \frac{H(r_s, \xi, t)}{\epsilon^c_\text{unif}(r_s, \xi)}
\end{equation}
\begin{figure}[!htb]
\includegraphics[width=\columnwidth,height=0.45\textwidth]{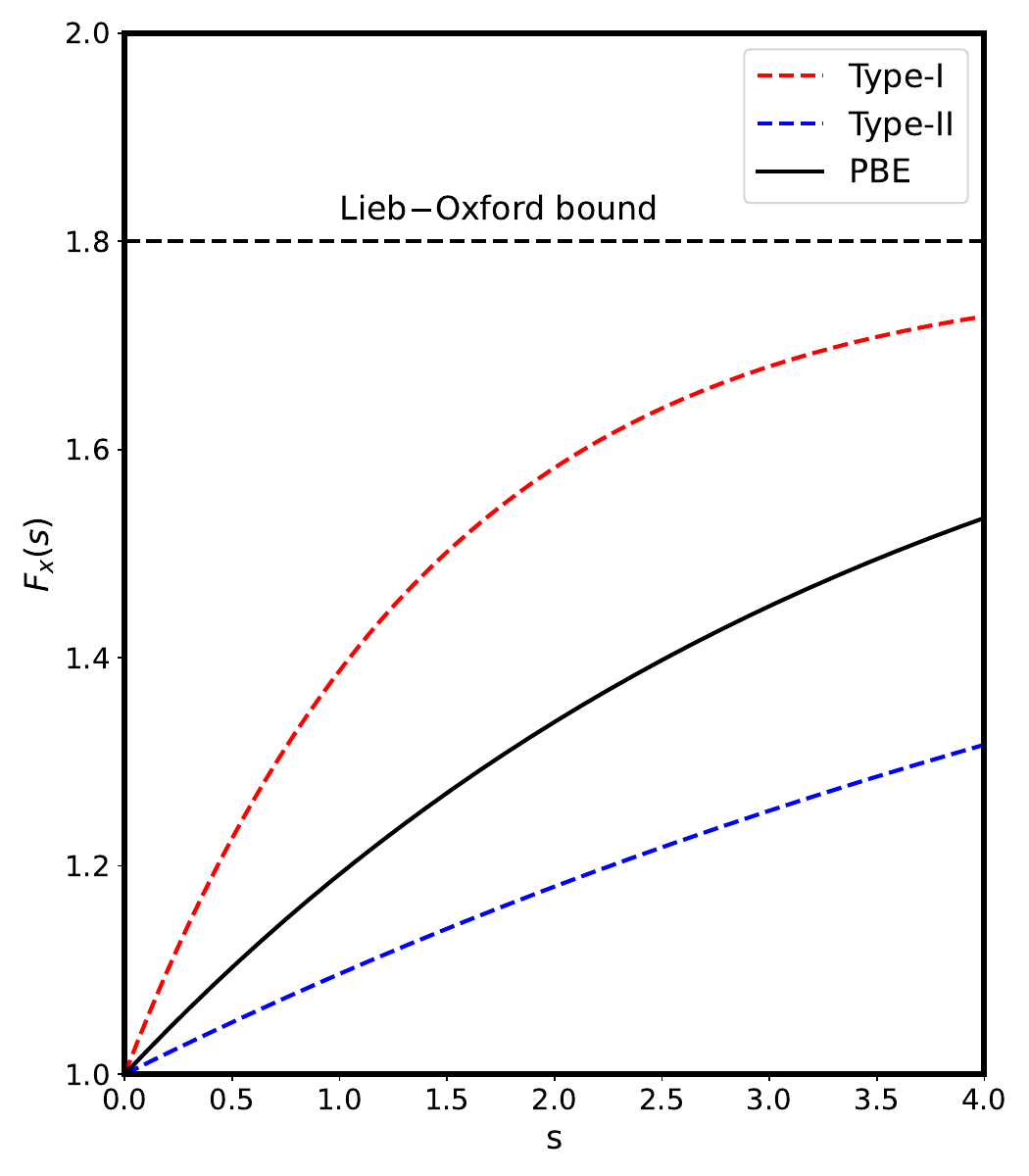}
\caption{Exchange enhancement factors $F_X$ versus density gradient for the two types of customized functionals for localized and delocalized cases respectively. For the PBE-functional (black curve), the curve is obtained from the standard definition as is given in Eq.\ref{X-en}. }
\label{fig:F_x}
\end{figure}
We propose a modified exchange enhancement factor shown below,
\begin{equation}
F^{present}_X(s)=1.0+\gamma\kappa\left (e^{-\alpha\mu_1s^2}-e^{-\beta\mu_1s^2}\right )
\end{equation}
We have introduced three more parameters $\alpha$, $\beta$ and $\gamma$ which can be used to further control the exchange-enhancement factor. $\mu_1=\frac{\mu}{\kappa}$. For type-I functional, we choose the following parameters:
$\alpha=0.01$, $\beta=2.5$ and $\gamma=0.98$ while for type-II we use $\alpha=0.01$, $\beta=0.5$ and $\gamma=0.96$. 
The new exchange enhancement factors introduced in our study contain three adjustable parameters: $\alpha$, $\beta$, and $\gamma$. 
The introduction of these parameters enables a more refined adjustment of the exchange energy in response to changes in the electron density gradient (\( s \)), potentially allowing the functional to approximate the properties such as band gap better. Also, this tailored approach facilitates the functional's application to a wider range of materials, including but not limited to Group III-V and IV-VI semiconductors for which we have demonstrated the results.
In Fig.\ref{fig:F_x}, we show the exchange enhancement  (dashed red \& blue curves) as a function of density gradient $s$ for two cases: The dashed red curve corresponds to type-I while the dashed blue curve corresponds to type-II functional. For comparison, we also plot the exchange enhancement factor $F^{\text{PBE}}_X(s)$, for the PBE functional (black curve). For all the cases, the local Lieb-Oxford bound is satisfied, i,e $F^{\text{type-I or II}}_X(s)\le 1.804$. It can be seen that for all the values of the reduced density gradient, the exchange energies are enhanced for type-I functional, while for type-II functional, it is decreased with respect to the PBE functional.
The function $H$ is given by~\cite{perdew1996generalized},
\begin{equation}
H=(e^{2}/a_{0})\gamma'\phi^{3}ln\Bigr\lbrack1\ +\frac{\beta'}{\gamma'}\,t^{2}\Bigl\lbrack\frac{1\ +\ A t^{2}}{1\ +\ A t^{2}+\ A^{2}t^{4}}\Bigr\rbrack\Bigr\rbrack
\end{equation}
Here $\beta'=0.066 725$ and $\gamma'=0.031 091$. The factor A is given by $A=\frac{\beta'}{\gamma'}\left[\exp\{-\epsilon^{\mathrm{c}}_{\mathrm{unif}}\slash(\gamma'\phi^{3}e^{2}\slash{a_{0}})\}-1\right]^{-1}$

\begin{table}[htbp]
  \centering
  \caption{Theoretical band gaps of different semiconductors with different methods compared with experimental ones.
  B3: zinc blende structure, B1: rock salt structure and B4: wurtzite structure. For Zn-chalcogenides, the PBE band gap in this table are those obtained in the present work. Other PBE, LDA and experimental band gap values are taken from \cite{tran2009accurate, hinuma2017band, heyd2005energy, tran2007band, gopal2015improved}.}
  \label{tab:Gaps}
  \begin{tabular}{lcccccc}
    \toprule
    Material & LDA & PBE & Present & Exp (eV) \\
     & (eV) & (eV)& (eV) (type-I or II)& \\
    \midrule
    Si        & 0.47  & 0.55  & 1.76 (I)    & 1.17 \\
    GaAs-B3   & 0.3   & 0.4   & 1.07 (II)    & 1.52 \\
    AlP-B3    & 1.46  & 1.56  & 2.95 (I)    & 2.45 \\
    GaN-B3    & 1.63  & 1.55  & 2.04 (II)    & 3.3 \\
    BN-B3     & 4.39  & 4.46  & 5.88 (I)    & 6.25 \\
    SiC-B3    & 1.35  & 1.46  & 2.63  (I)   & 2.42 \\
    LiCl-B1   & 6.06  & 6.41  & 7.07 (II)    & 9.4 \\
    ScN-B1    & 0.0   & 0.0   & 0.67 (I)    & 0.9 \\
    MgO-B1    & 4.7   & 4.73  & 5.05 (II)    & 7.8 \\
    LiF-B1    & 8.94  & 9.19  & 9.89 (II)    & 14.3 \\
    AlN-B4    & 4.17  & 4.16  & 4.41 (I)    & 6.28 \\
    CdS-B3    & 0.86  & 1.11  & 1.73 (II)   & 2.55 \\
    ZnS-B4    & 2.45  & 2.07  & 3.25 (II)    & 3.86 \\
    ZnS-B3    & 1.84  & 2.01  & 2.9 (II)     & 3.66 \\
    ZnSe-B4   & 1.43  & 1.18  & 2.43 (II)    & 2.7 \\
    ZnSe-B3   & 1.21  & 1.14  & 2.03 (II)    & 2.82 \\
    ZnO-B4    & 0.75  & 0.72  & 1.37 (I)     & 3.4 \\
    ZnO-B3    & 0.71  & 0.62  & 1.31 (I)    & 3.27 \\
    \bottomrule
  \end{tabular}
\end{table}
The variable $t$ in the PBE functional is a measure of the gradient of the electron density, which reflects the variation of the density in space. The value of $t$ depends on the shape and size of the orbitals that contribute to the density. For example, s orbitals are spherical and have no nodes,  On the other hand, p orbitals are directional and have one node, therefore the values of $t$ will be quite different in the two respective cases. Similarly, d orbitals have even more nodes and directions, so they have an even larger gradient and value of $t$. Therefore, the variable \(t\) is different for s, p, and d-orbitals.
For spin-unpolarized systems, where \( \zeta = 0 \), \( \phi \) is set to 1 and the parameter t reduces to $t = \frac{|\nabla n|}{2k_s n}$. The parameter \( t \) essentially measures the relative importance of the gradient of the electron density \( |\nabla n| \) compared to the local density \( n \) itself, scaled by the Thomas-Fermi screening wave vector \( k_s \).
In this work, we rescale the parameter $t$. For type-I, we chose $t$ to be scaled by a factor of 2.5 to consider a larger electron correlation effect while for the case of type-II, the scaling factor we choose is 0.5. These choices for the rescaling factor are motivated by the fact that second-order gradient expansion term for exchange cancels the corresponding correlation term as was initially proposed in the construction of PBE-functional. To reduce $t$, we essentially reduce the electron density gradient ($\nabla n$) or increase the  Thomas-Fermi screening wave number ($k_s$). Reducing the gradient of the electron density ($\nabla n$) typically means the electron density is becoming more uniform across the material. A more uniform electron density usually indicates less localization, as electrons are more evenly distributed over space. Also, increasing $k_s$ means larger screening. Stronger screening could lead to more delocalized electrons since the external potentials (which might localize electrons) are more effectively screened. Therefore instead of delocalizing electrons with a negative U as is done in the DFT+U approach we rescale the reduced density gradient  \( t \) further.

It should be noted here, for the present study, we have exclusively implemented these customized functionals for the non-magnetic semiconductors only ($\phi=1$), since scaling of $t$ ($= \frac{|\nabla n|}{(2\phi k_s n)}$) in magnetic systems introduces a complex spin scaling factor, $\phi = [\left(1 + \zeta\right)^{2/3} + \left(1 - \zeta\right)^{2/3}]/2$, which has resulted in inaccuracies in predicting magnetic moments.
\begin{figure}[!htb]
\includegraphics[width=\columnwidth,height=0.45\textwidth]{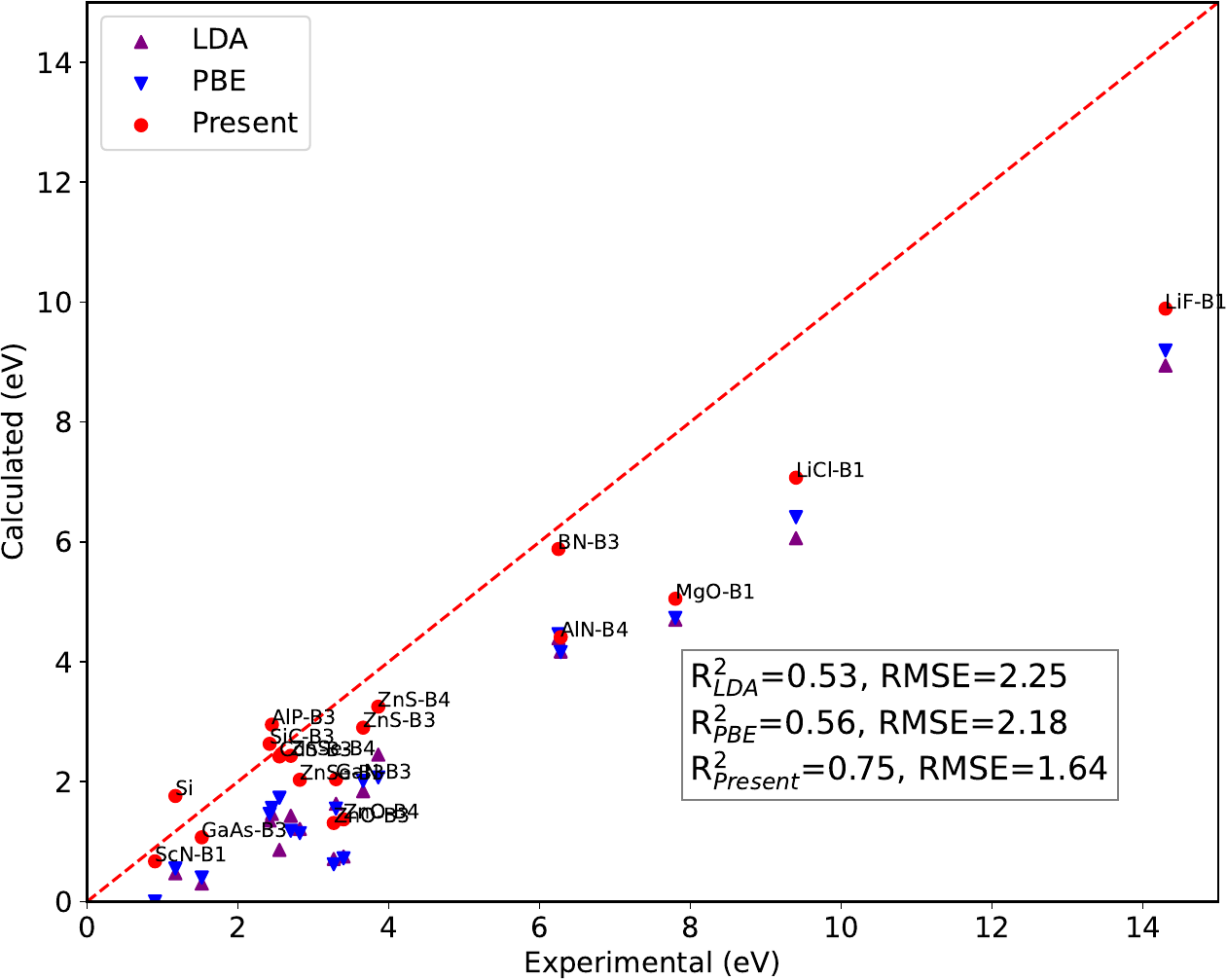}
\caption{Calculated  versus experimental band
gaps. The errors are also shown in the inset.}
\label{fig:Corr}
\end{figure}
\par
We have implemented the above modifications in the \textit{Vienna Ab initio Simulation Package} (\verb|VASP|) code~\cite{Kresse1993ab} which was used for further calculations. The interaction between electrons and atomic cores is described by pseudopotentials generated in accordance with the Projector Augmented Wave (PAW) method \cite{Blochl1994}. The electron wavefunctions are expanded on plane waves with energy cutoff of 500 eV. All atomic positions and lattice parameters are relaxed such that the residual force on any atom is less than 0.01 eV/\AA. The total energy per cell converged below 10$^{-6}$ eV.
\section{The choice of the parameters and linear response constraint}
The selection of the parameters is strategically motivated by the $s\rightarrow{0}$ behavior of the exchange enhancement factors, aiming for flexibility across a diverse class of materials as well as to meet the linear response constraint. Our approach, by integrating these parameters, is fundamentally designed to modulate the PBE exchange enhancement factor in a controlled manner. This modulation is evident in the $s\rightarrow{0}$ limit, where the original PBE form, $F_{X}^{\mathrm{PBE}}(s) = 1 + \mu s^{2}$, is transformed into $F_{X}^{\mathrm{Present}}(s) = 1 + [\gamma(\beta-\alpha)]\mu s^{2}$. Here, $\alpha$ is maintained at a constant value of 0.01, serving as a baseline for both upscaling and downscaling scenarios. The adjustment of $\gamma$ is minor, whereas $\beta$ is the key variable, manipulated from 2.5 (to upscale) to 0.5 (to downscale) the exchange enhancement effect as required. It is important to note that, the introduction of the new parameters, however, maintains the linear response constraint. As we can see from the above, in the slowly-varying limit ($s\rightarrow{0}$) for the type-I functional, the factor $\gamma(\beta-\alpha)\sim 2.5$ is same as the scaling factor for the $t$, similarly for the type-II functional, the $\gamma(\beta-\alpha)\sim 0.5$ is almost same as the scaling factor for the $t$. Therefore, the second-order gradient expansion term for exchange cancels the corresponding correlation term something which is needed to maintain the linear response constraint.
\section{Results and discussion}
The Table-\ref{tab:Gaps} presents a comparison of the band gaps of various semiconductors calculated using different theoretical methods LDA, PBE and newly proposed functionals referred to as type-I and type-II, alongside their experimental values (Exp). The traditional LDA and PBE functionals generally underestimate the band gaps compared to experimental values, which is a well-known limitation of these approximations. The type-I and type-II functionals introduced for the calculation of semiconductor band gaps address the inherent deficiencies in the LDA and PBE functionals related to exchange and electron correlation effects. Type-I functional is particularly useful in situations where electron localization is crucial and where LDA and PBE tend to underestimate these effects. Conversely, type-II functional is beneficial when electron localization is overestimated by LDA and PBE, as they enhance electron-correlation and exchange enhancement effects. The semiconductors ZnO, ZnS, and ZnSe serve as illustrative examples. For ZnO, which exhibits a strong electron correlation as well as an exchange enhancement effect, PBE presents a smaller band gap that is modestly improved upon by LDA. However, type-I functional, which boosts the electron correlation and exchange enhancement, yields a more robust and accurate band gap. In contrast, for ZnS and ZnSe, the weak correlation effects and exchange enhancement factor are better accounted for by LDA (compared to PBE), but it is the type-II functional that significantly reduces the electron correlation effect, thereby providing a more precise band gap estimation closer to experimental values. This illustrates the effectiveness of these functionals in cases where either a reduction in or an enhancement of electron correlation is needed to align with observed experimental data. 
The Fig.\ref{fig:Corr} displays a scatter plot comparing calculated band gaps using LDA, PBE, and the currently discussed \textit{Present} functional (either type-I or type-II) against the experimental band gap values for a range of semiconductors. The dotted red line represents perfect agreement between calculated and experimental values. The calculated values of the band gap are evaluated via the metrics, R-squared ($R^2$) and Root Mean Square Error (RMSE) in comparison to the actual experimental values.
The figure indicates that the present approach (type-I and type-II) achieves the best statistical correlation with the experimental band gap values among the three theoretical approaches. The R$^2$ value for the present approach is 0.75, indicating a strong positive correlation between the calculated and experimental values, and this is substantially better than the R$^2$ values for LDA and PBE, which are 0.53 and 0.56, respectively. The RMSE, which measures the square root of the average squared error between predicted and observed values, is lowest for the present approach at 1.64 eV. This suggests that the present approach not only predicts band gap values that are, on average, closer to the experimental values but also that its predictions are more consistent. 
\subsection{Evaluating consistency and accuracy in band Gap calculations across contemporary density functionals}
In this subsection we compare the band gaps obtained using the present functional with those obtained with SCAN (Strongly Constrained and Appropriately Normed semilocal density functional)~\cite{SCAN} and HSE functional. For the sake of the comparison we consider the subset of the materials in the table-\ref{tab:Gaps} for which all band gap values for all the three funcionals are available. The band gap values for the HSE  and SCAN functional are taken from the multiple sources ~\cite{heyd2005energy,patra2019efficient,gopal2015improved} while for the SCAN functional we have used only the result reported in Patra \textit{et al}~\cite{patra2019efficient}.
The correlation between the experimental and calculated band gaps is quantified by the coefficient of determination $R^2$ and the root-mean-square error (RMSE). The $R^2$ values indicate that the HSE functional has the highest correlation with the experimental data ($R^2_{\text{HSE}} = 0.94$), followed by the SCAN functional ($R^2_{\text{SCAN}} = 0.77$). The present functional has very similar $R^2$ value ($R^2_{\text{Present}} = 0.76$). 
\begin{figure}[!htb]
\includegraphics[width=\columnwidth,height=0.5\textwidth]{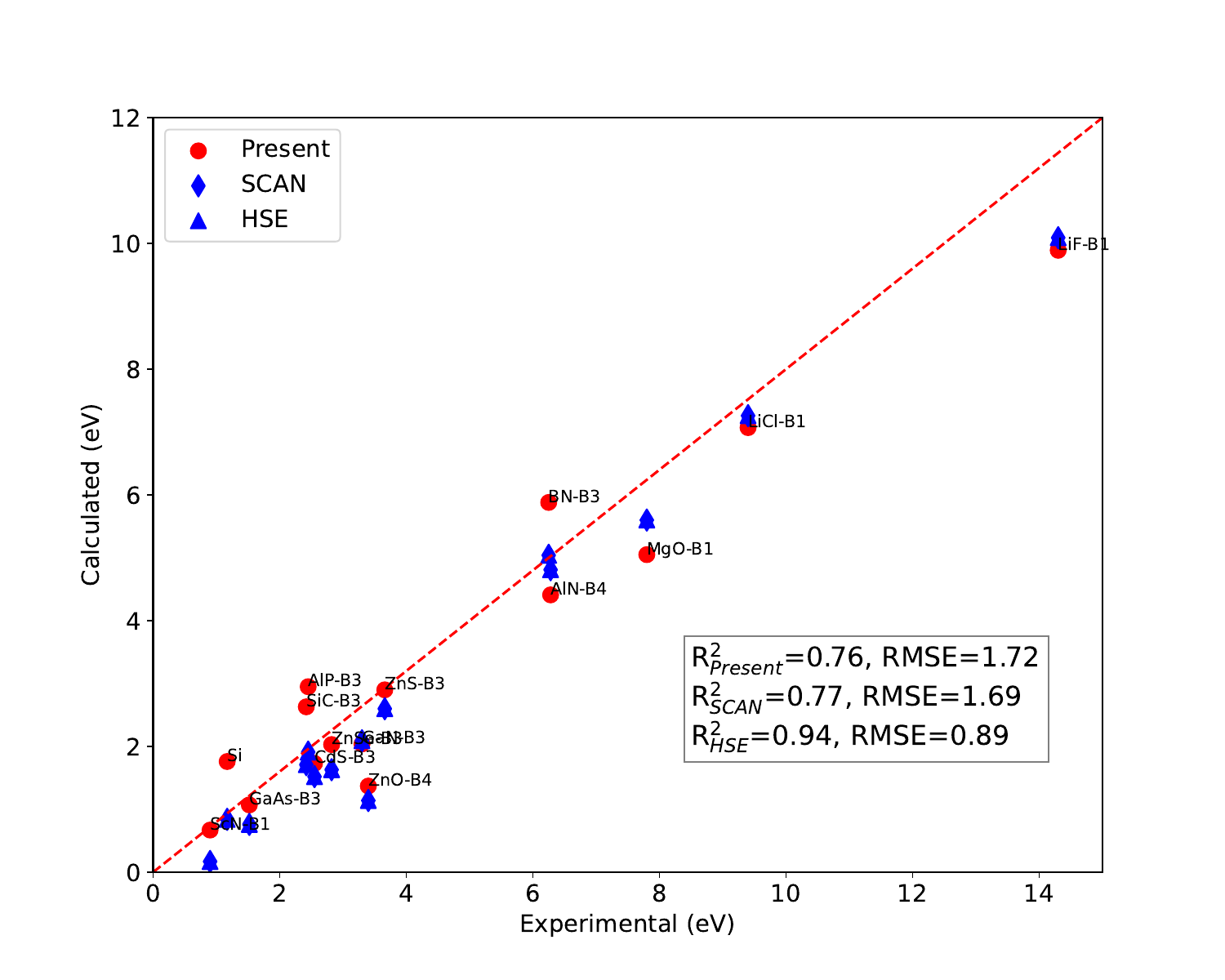}
\caption{Comparison of the band gaps obtained using Present functional with SCAN and HSE functional.}
\label{fig:Comp}
\end{figure}
As expected in terms of predictive accuracy, the HSE functional also demonstrates the lowest RMSE ($\text{RMSE}_{\text{HSE}} = 0.89$ eV), suggesting it has a superior predictive capability for the band gaps in comparison to the present functional ($\text{RMSE}_{\text{Present}} = 1.72$ eV) and the SCAN functional ($\text{RMSE}_{\text{SCAN}} = 1.69$ eV). However from the point of view of coefficients of determination ($R^2$) and the RMSE values, which provide a measure of the differences between values predicted by a model and the values actually observed, the Present funcional is very similar to the SCAN functional.
\subsection{Lattice constants, effective masses, dielectric constants and Born effective charges}
In Table-\ref{Lattice}, we compare the lattice constants of various semiconducting materials calculated using the PBE and the present functionals against experimental values. Overall, the present approach tends to give lattice constants closer to the experimental values than the PBE calculations. This is particularly evident for materials with B3 and B1 structures, where the present approach corrects the underestimation of lattice parameters typically observed in PBE calculations. For instance, for GaN-B3, the present method estimates the lattice constants at 4.48\AA~which is much closer to the experimental range of 4.52-4.53\AA~compared to the PBE's 4.55, 4.57\AA. However, for B4 structured materials like ZnO, the present approach yields 3.45, 5.57 \AA,~ whereas PBE’s 3.28, 5.31 \AA~ align better with the experimental measures of 3.26, 5.22 \AA. These results suggest that the better predictions for the band gaps come with minor adjustments in the case of lattice constants. 
Finally, to further showcase our findings, we highlight the results obtained for the electron effective mass and electronic dielectric constant of Zn-monochalcogenides, calculated using both the standard PBE and the customized functionals. The values are tabulated in the Table-\ref{EM}\& \ref{DC}. It is found that the effective masses obtained
using customized functionals match better with the experimental data when compared to standard PBE functional. For example, for the \( \Gamma \rightarrow X \) transition in zinc blende structure, the customized functional shows improved predictions, such as in ZnSe, where the value is corrected from \(0.12\) to \(0.16\), closer to the experimental reference of \(0.15\). 
For ZnO in the wurtzite structure, the effective mass along \( K \rightarrow \Gamma \) is negative (-0.65), while it is positive (0.20) along \( \Gamma \rightarrow A \). This significant asymmetry in effective mass for different crystallographic directions implies a pronounced anisotropic behavior in its electronic structure, which will manifest in its transport properties as direction-dependent electron mobility.
In contrast, for ZnS and ZnSe in both wurtzite and zinc blende phases, the effective masses in the \( L \rightarrow \Gamma \) and \( \Gamma \rightarrow X \) directions are indeed closer in value, indicating a more isotropic behavior in comparison to ZnO.
These variations in effective mass anisotropy among ZnO, ZnS, and ZnSe will influence their electronic and optical properties, as well as the efficiency of processes such as charge carrier transport and recombination, which are critical in applications like photocatalysis and optoelectronics.
The Table-\ref{DC} shows the calculated dielectric constant, \(\epsilon_{\infty}\), for various compounds in two different crystal structures: wurtzite and zinc blende. The calculations have been performed using two different approaches: customized functionals and the PBE functional. The dielectric constant \(\epsilon_{\infty}\) values were calculated using density functional perturbation theory (DFPT)~\cite{gajdovs2006linear}.

In both the wurtzite and zinc blende structures, the customized functionals appear to yield calculated values of \(\epsilon_{\infty}\) that are generally closer to the experimental data compared to the values calculated using the PBE functional. This indicates a better performance of the customized functionals over the standard PBE approach for these materials.

For ZnO, the PBE significantly underestimates the dielectric constant compared to the experimental value in both structures, but the discrepancy is more pronounced in the wurtzite structure. The customized functional corrects this, bringing the calculated value much closer to the experimental data, although there's still a noticeable difference.

For ZnS, the custom functional and the PBE results are almost identical in the wurtzite structure and are both in good agreement with the experiment. In the zinc blende structure, the customized functional slightly improves upon the PBE result when compared to the experimental value.

For ZnSe, the customized functionals again provide a closer approximation to the experimental data in both structures, with the PBE result significantly underestimating the dielectric constant, particularly in the zinc blende structure.

The improvement seen with the customized functionals suggests that these functionals are better at capturing the exchange-correlation effects that influence the electronic structure, and thus the dielectric properties, of these materials. Since the electronic dielectric constant is related to the polarizability of the ion cores and the electronic cloud around them, a more accurate functional can capture the response of these electrons to an external electric field more effectively. These findings further suggest that the customized functionals we presented will be suitable for \textit{ab-initio} transport calculations of these materials without requiring a substantial amount of computational resources. 
\begin{table}[h]
\centering
\begin{tabular}{|l|l|l|l|}
\hline
\textbf{Material} & \textbf{PBE (\AA)} & \textbf{Present } & \textbf{Exp (\AA)} \\ \hline
Si                & 5.48               & 5.62                                      & 5.42               \\ \hline
GaAs-B3           & 5.76               & 5.56                                      & 5.64, 5.65         \\ \hline
AlP-B3            & 5.51               & 5.63                                      & 5.45, 5.46         \\ \hline
GaN-B3            & 4.55, 4.57         & 4.48                                      & 4.52, 4.53         \\ \hline
BN-B3             & 3.63               & 3.66                                      & 3.59, 3.61         \\ \hline
SiC-B3            & 4.38, 4.40         & 4.46                                      & 4.34, 4.36         \\ \hline
LiCl-B1           & 5.17               & 4.95                                      & 5.07, 5.11         \\ \hline
ScN-B1            & 4.51 (PW-GGA)      & 4.64                                      & 4.51               \\ \hline
MgO-B1            & 4.26               & 4.17                                      & 4.19, 4.21         \\ \hline
LiF-B1            & 4.07               & 3.94                                      & 3.96, 4.01         \\ \hline
AlN-B4            & a=3.13, c= 5.04    & a=3.17, c=5.12                            & a=3.11, c=4.98     \\ \hline
CdS-B3            & 5.96         & 5.68                                      & 5.82               \\ \hline
ZnS-B4            & a=3.88, c=6.30     & a=3.59, c=5.90                            & a=3.81, c=6.23     \\ \hline
ZnS-B3            & 5.45         & 2.90                                      & 5.41               \\ \hline
ZnSe-B4           & a=4.04, c=6.70     & a=3.76, c=6.17                            & a=3.99, c=6.53     \\ \hline
ZnSe-B3           & 5.74        & 5.46                                      & 5.67               \\ \hline
ZnO-B4            & a=3.28, c=5.31     & a=3.45, c=5.57                            & a=3.26, c=5.22     \\ \hline
ZnO-B3            & 4.63               & 4.85                                      & 4.67              \\ \hline
\end{tabular}
\caption{A comparison of lattice constants obtained via PBE and present approach. B1 and B3 structures correspond to the cubic ones while B4 corresponds to wurtzite. For a better comparison, the experimental values are also shown.~\cite{friak2018ab,haas2009calculation,moses2011hybrid,gopal2015improved}.}
\label{Lattice}
\end{table}

\begin{table}[h]
\centering
\begin{tabular}{|c|cc|c|}
\hline
\multicolumn{1}{|c|}{\multirow{2}{*}{}} & \multicolumn{2}{c|}{Wurtzite} & \multirow{2}{*}{Exp. data} \\
\multicolumn{1}{|c|}{}                  & K $\rightarrow$ $\Gamma$       & $\Gamma$ $\rightarrow$ A       &                        \\ \hline
ZnO                                     & -0.65 (-0.44)           & 0.20 (0.17)             & 0.21 ($\|$) ~\cite{Shokhovets2006}         \\
                                        &                         &                         & 0.24 ($\perp$) [~\cite{Shokhovets2006}    \\
\hline
ZnS                                     & 0.56 (0.63)             & 0.21 (0.18)            & 0.28 ~\cite{Miklosz1967}              \\
ZnSe                                    & 0.51 (0.58)             & 0.15 (0.12)             & -                      \\ \hline
\multicolumn{1}{|c|}{\multirow{2}{*}{}} & \multicolumn{2}{c|}{Zinc blende} & \multirow{2}{*}{Exp. data} \\
\multicolumn{1}{|c|}{}                  & L $\rightarrow$ $\Gamma$      & $\Gamma$ $\rightarrow$ X       &                        \\ \hline
ZnO                                     & -0.40 (-0.39)           & 0.20 (0.17)             & -                      \\
ZnS                                     & 2.53 (2.84)             & 0.23 (0.19)             & 0.22 ~\cite{Shokhovets2007}              \\
ZnSe                                    & 2.31 (2.51)             & 0.16 (0.12)             & 0.15 ~\cite{Shokhovets2007}             \\ \hline
\end{tabular}
\caption{The calculated electron effective masses using the customized PBE-functional compared with experimental data. The ones with traditional PBE functional are shown in brackets.}
\label{EM}
\end{table}
\begin{table}[h]
\centering
\begin{tabular}{|c|c|c|c|c|c|c|c|}
\hline & \multicolumn{2}{|c|}{ Wurtzite } & Exp. data & \multicolumn{2}{c|}{ Zinc blende } & Exp. data \\
\hline & Customized & PBE & & Customized & PBE & \\
\hline ZnO & 4.87 & 6.41 & $3.92$~\cite{Goni2014} & 4.75 & 6.12 & \\
\hline ZnS & 5.78 & 5.88 & $5.7$~\cite{Manabe1967} & 5.68 & 5.91 & $5.7$~\cite{Manabe1967} \\
\hline ZnSe & 6.65 & 7.50 & & 6.62 & 7.40 & $5.4$~\cite{Manabe1967} \\
\hline
\end{tabular}
\caption{Calculated dielectric constant ($\epsilon_\infty$) using the customized functionals and PBE approach.}
\label{DC}
\end{table}
\subsection{Electronic density of states of Zn-monochalcogenides}
Let us now look at the density of states (DOS) profiles of Zn-monochalcogenides as model systems for showcasing our proposed functionals.  As mentioned, Zn-monochalcogenides materials allow us to showcase the results for type-I and type-II functionals. 
\par
In the case of ZnO, the main feature of the valence band edge is mainly defined by oxygen \(p\) states, coupled with a notable presence of Zn \(3d\) orbitals, showing the profound hybridization between Zn and O states. One important factor is the position of the d-band, which is crucial in determining the electronic structure of materials, influencing their optical and electrical properties. The d-band position is defined in terms of the average energy of the Zn-3d band, which is the center of gravity of the occupied part of the Zn-3d band. Note that the position of the average energy of the Zn-3d band is at around -4.4 eV, which is slightly higher than PBE (-4.8 eV) reported by Wrobel \textit{et al.}~\cite{wrobel2009calculations}. The same study also found that the Zn-d average energy for the HSE method is around -6 eV. However, all of these studies have reported higher values than the experimentally obtained value (-7.5 to -7.8 eV) through photoemission experiments~\cite{reynolds1999valence}. It can be seen that the average d-band energy for the ZnO for both zinc blende and wurtzite lie at the same energy. This situation is slightly different for ZnS where the average d-band energy of the Wurtzite phase (-6.3 eV) is slightly pushed down with respect to the corresponding value for the zinc blende phase (-5.8 eV). A similar trend is also observed for ZnSe where for wurtzite phase the average d-band energy (-6.6 eV) is lower in comparison to that of the Zinc Blende phase (-6.3 eV).
\begin{figure}[!htb]
\includegraphics[width=1.0\columnwidth,height=0.42\textwidth]{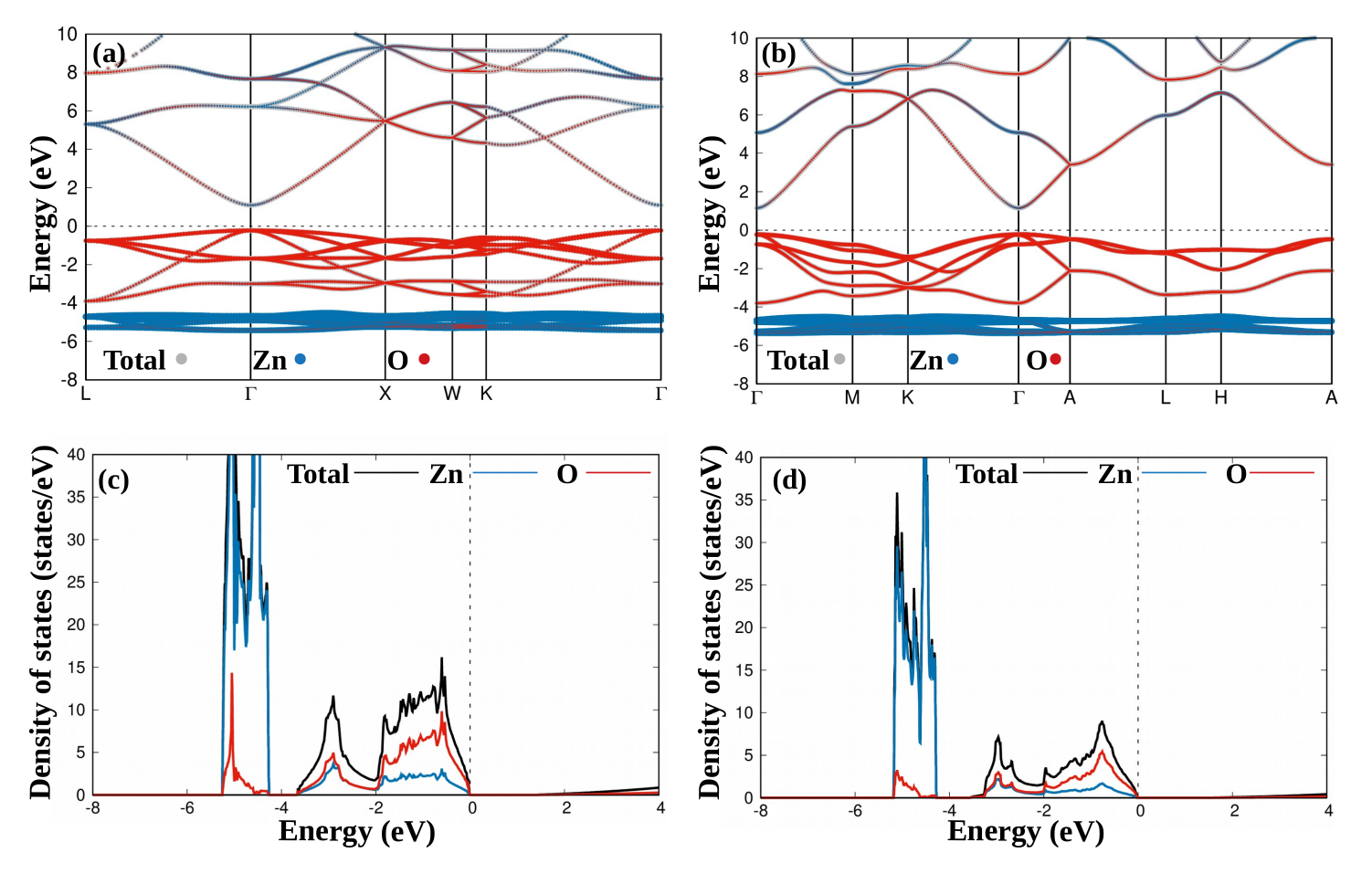}
\caption{Electronic Band structure and Density of states for ZnO using the customized PBE functional. Left panels( (a)\&(c)): zinc blende, right panels ( (b)\&(d)): wurtzite}
\label{fig:enter-label}
\end{figure}
\begin{figure}[!htb]
\includegraphics[width=1.0\columnwidth,height=0.42\textwidth]{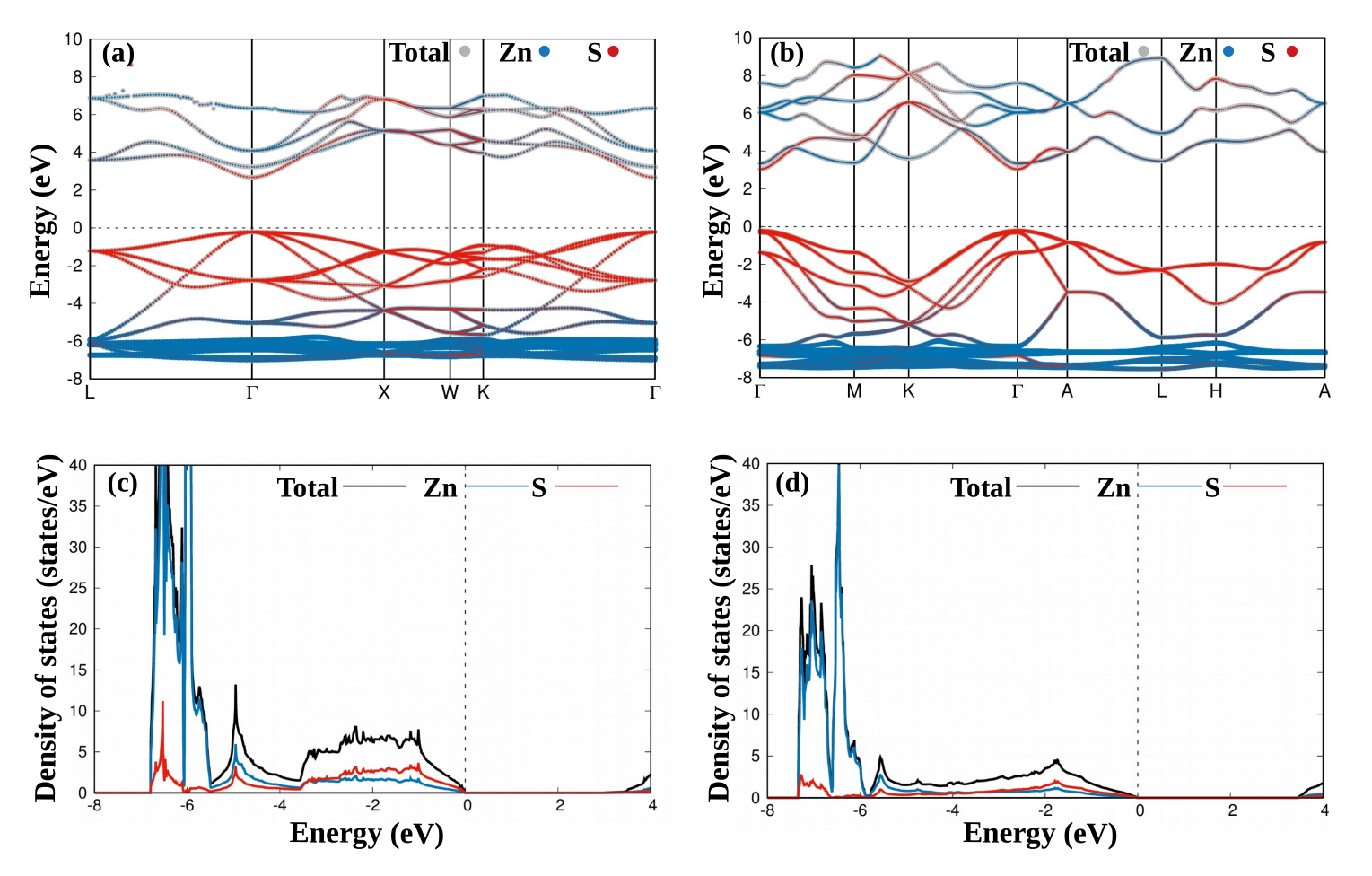}
\caption{Electronic Band structure and Density of states for ZnS using the customized PBE functional. Left panels( (a)\&(c)): zinc blende, right panels ( (b)\&(d)): wurtzite}
\label{fig:enter-label}
\end{figure}
\begin{figure}[!htb]
\includegraphics[width=1.0\columnwidth,height=0.42\textwidth]{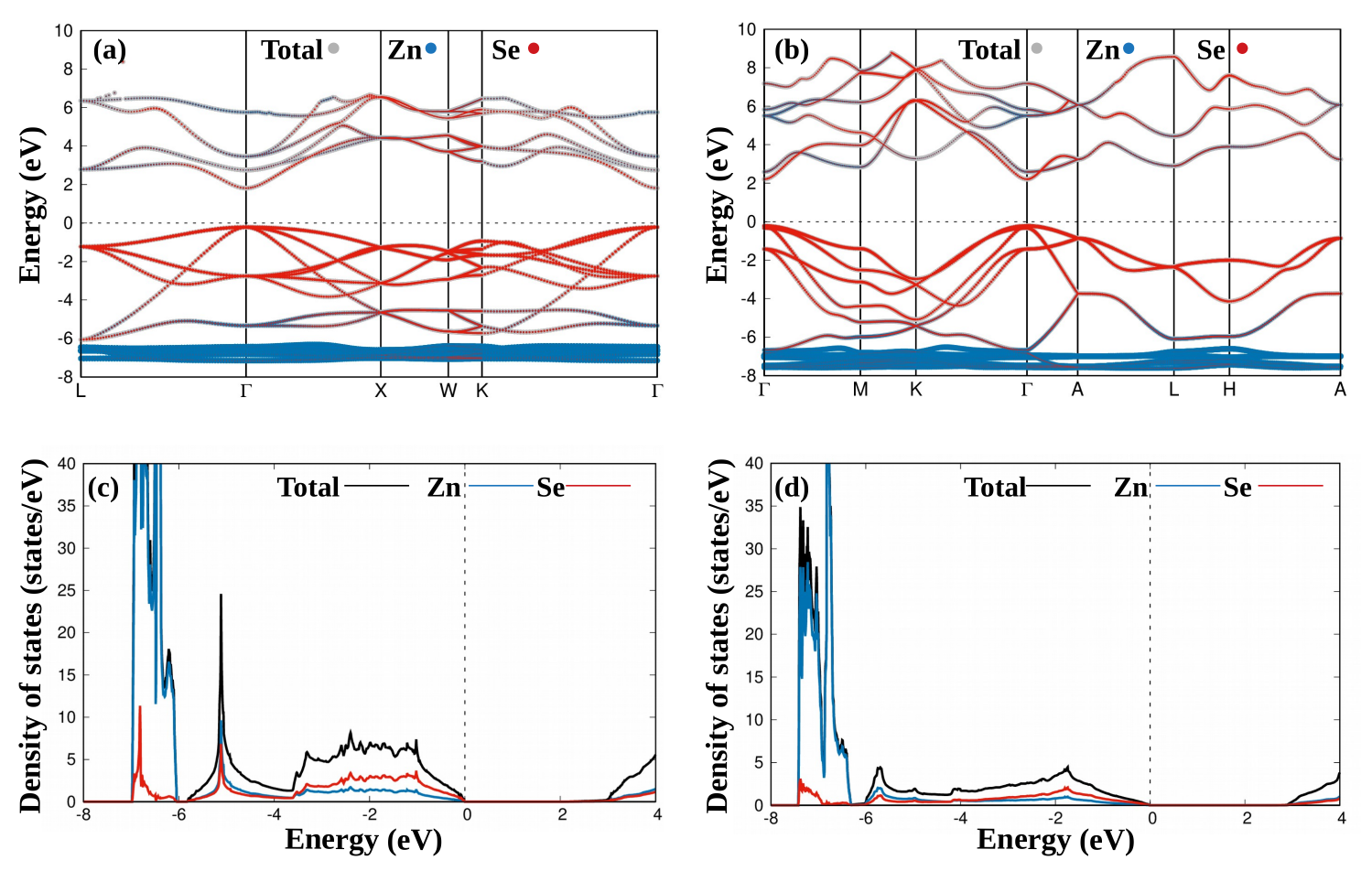}
\caption{Electronic Band structure and Density of states for ZnSe using the customized functional. Left panels( (a)\&(c)): zinc blende, right panels ( (b)\&(d)): wurtzite}
\label{fig:enter-label}
\end{figure}
It can be seen that the energy window of Zn-chalcogen hybridization in the valence band region increases from ZnO to ZnSe. Furthermore, as seen in the literature, all these compounds are direct band semiconductors~\cite{karazhanov2006coulomb} is also consistent with the present study. 
Therefore, in overall the comparative analysis across these semiconductors illustrates the profound impact of our customized exchange-correlation functionals on the theoretical understanding of electronic structures. By surpassing the limitations of the PBE functional, they allow for a more accurate and comprehensive investigation into the orbital contributions that define the electronic and optical properties of these semiconducting systems. 

\section{Selection of type-I and type-II functionals based on simple atomic features}
\begin{figure}[h]
\includegraphics[width=1.1\columnwidth]{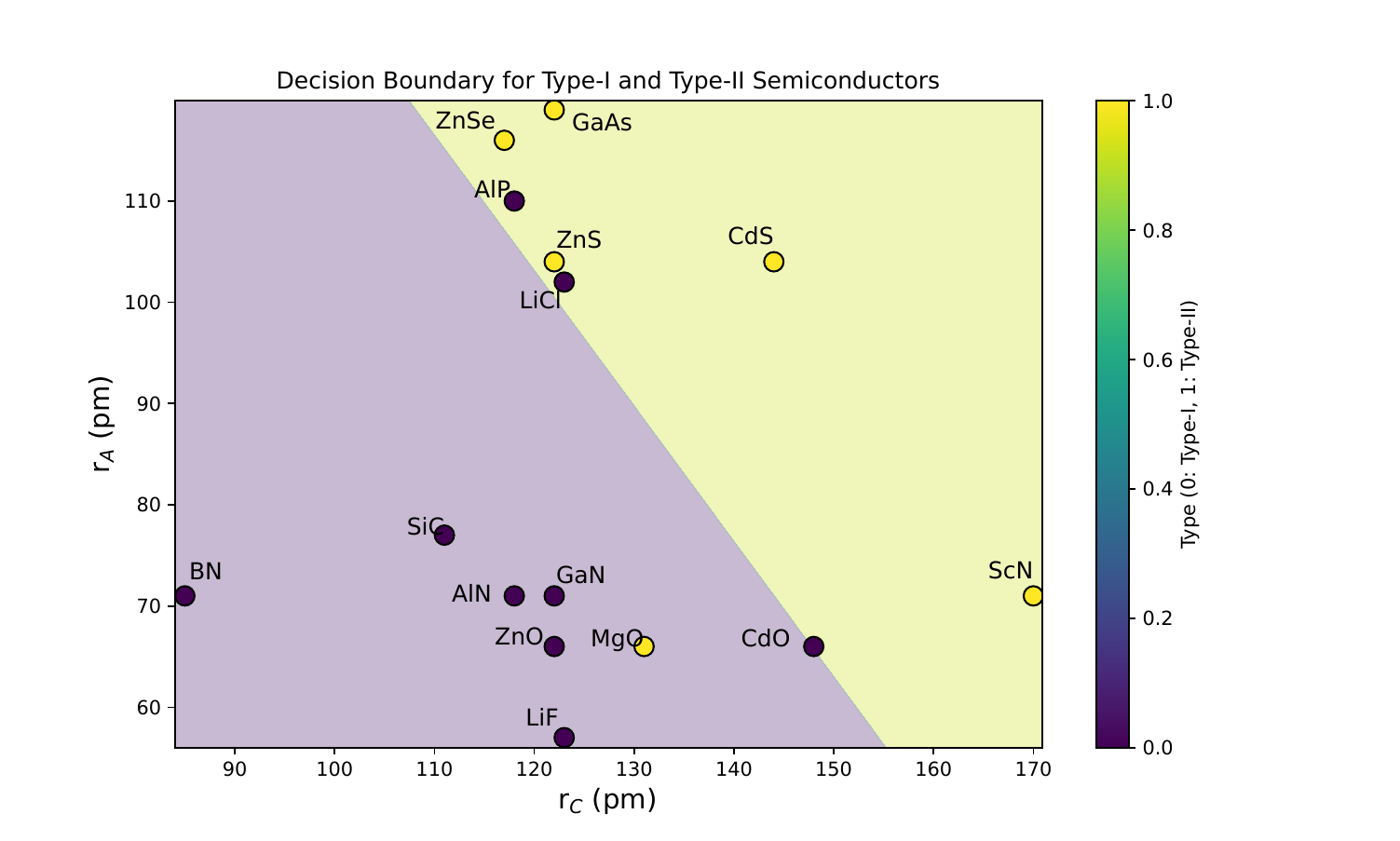}
\caption{K-means clustering results of semiconductor materials in the \((r_C, r_A)\) space with decision boundary obtained via a logistic regression.}
\label{Clustering}
\end{figure}

In this final section, we discuss the suitability of a given functional for a particular semiconducting material. For this, we performed a classification task using simple elemental features as described in the following. We start with a dataset of the materials as reported in the table-\ref{tab:Gaps}, each characterized by two simple features: cation radius ($r_C$) and anion radius ($r_A$). The materials are categorized as type-I or type-II based on their band gap properties based on the functional that best describes their band gap. With this approach, for example, ZnO can be termed as type-I material and ZnS can be termed as type-II material. Since Si is not a binary material, we drop it from the database, and instead, we include CdO (B4) which from our calculation shows a type-I behavior with a band gap of 0.48 eV (PBE band gap is zero~\cite{friak2018ab}).
For the features, we use the data shown in the table-\ref{features}. Specifically, we focus on clustering the materials in the \((r_C, r_A)\) space and subsequently drawing a decision boundary to separate the two types. The covalent radii of the cation \((r_C)\) and the anion \((r_A)\) serve as the primary features for this analysis. A logistic regression model is employed to determine the decision boundary, and its accuracy in classifying the materials is also evaluated.
\begin{table}[h]
\centering
\caption{Data Preparation Table for Semiconductor Materials}
\begin{tabular}{|c|c|c|c|}
\hline
Material & $r_C$ (pm) & $r_A$ (pm) & Type \\
\hline
GaAs  & 122 & 119 & II \\
AlP   & 118 & 110 & I  \\
GaN   & 122 & 71  & I  \\
BN    & 85  & 71  & I  \\
SiC   & 111 & 77  & I  \\
LiCl  & 123 & 102 & I  \\
ScN   & 170 & 71  & II \\
MgO   & 131 & 66  & II \\
LiF   & 123 & 57  & I  \\
AlN   & 118 & 71  & I  \\
CdS   & 144 & 104 & II \\
ZnS   & 122 & 104 & II \\
ZnSe  & 117 & 116 & II \\
ZnO   & 122 & 66  & I  \\
CdO   & 148 & 66  & I  \\
\hline
\end{tabular}
\label{features}
\end{table}
To determine if the type-I and type-II materials form distinct clusters in the \((r_C, r_A)\) space, we applied the K-means clustering algorithm. K-means clustering partitions the data into \(k\) clusters by minimizing the variance within each cluster. In this case, we set \(k = 2\) to reflect the two types of materials.
The clustering results, illustrated in Fig.\ref{Clustering}, show two clusters represented by different colors. Each data point is labeled with the corresponding material name for clarity.
To draw a decision boundary that separates type-I and type-II materials, we employed a logistic regression model. Logistic regression is used for binary classification that predicts the probability of a binary outcome based on one or more predictor variables. Within this approach, we model the probability \( P(\text{Type-I or type-II}  = 1 | \mathbf{X}) \) that either of the binary response variable \( \text{Type-I/Type-II} \) is equal to 1 given the predictor variables \( \mathbf{X} = (r_C, r_A) \). The decision boundary is defined as the set of points where this probability is 0.5.

The logistic regression model was trained using the \((r_C, r_A)\) values as input features and the material types (type-I as 0 and type-II as 1) as labels.
The accuracy of the logistic regression model was evaluated by comparing the predicted types to the actual types in the dataset. We used the entire dataset. The model achieved an accuracy of approximately 73.33\%, indicating a reasonable performance given the limited feature set.
\[
\text{Accuracy} = \frac{\text{Number of correct predictions}}{\text{Total number of predictions}} = 73.33\%
\]
The classification of semiconductors into type-I and type-II can be quite well described by the radii of the cations and anions. From the above analysis it can be seen that materials with higher anion radii belong to type-II semiconductors (whose band gap can be calculated using type-II functional). On the other hand, when the anion radii are smaller and more comparable to the cation radii, the material is more likely to exhibit type-I behavior. Thus among ZnO and ZnS, due to the difference in the anion size, ZnO is well described by the type-I functional while ZnS is well described by the type-II functional. The same applies to CdO and CdS, which is clearly visible in Fig.\ref{Clustering}.
\section{Conclusion}
In conclusion, this study introduces two customized PBE functionals, designated as type-I and type-II, to effectively rectify the limitations observed in LDA and standard PBE functionals regarding the prediction of semiconductor band gaps.  These functionals modify the exchange enhancement factor and correlation effects, aligning them more closely with experimental data. Type-I functional is particularly effective for materials with strong electron correlations and exchange enhancement, such as ZnO where standard PBE functional underestimates the exchange-correlation hole, while type-II functional excels in materials like ZnS and ZnSe, where standard PBE functional overestimates the exchange-correlation hole. The proposed functionals provide more accurate band gap predictions and improve the calculation of electron-effective mass and dielectric constants with minimum adjustments to the lattice constants. These results indicate that the customized functionals offer a practical and efficient alternative for accurate \textit{ab-initio} transport calculations in semiconductors, without extensive computational resources as is needed for sophisticated approaches such as HSE, GW and meta-GGA~\cite{aschebrock2019ultranonlocality}.  Consequently, this paves the way for the seamless integration of these functionals into workflows that begin with basic DFT calculations and extend to sophisticated post-processing for diverse applications, ranging from the study of transport to optical properties.

\section{Acknowledgment}
This work was supported by the Korea Institute of Science and Technology (Grant number 2E31851), GKP (Global Knowledge Platform, Grant number 2V6760) project of the Ministry of Science, ICT and Future Planning.
\section{Competing interests}
The authors declare no competing financial interests.
\bibliography{Ref}
\bibliographystyle{apsrev4-2}
\end{document}